\documentstyle[12pt,epsf]{article}
\begin{document}
\begin{titlepage}
\pagestyle{empty}
\baselineskip=21pt
\rightline{Alberta Thy-28-95}
\rightline{December 1995}
\vskip .2in
\begin{center}
{\large{\bf Nucleon Decay in Non-Minimal Supersymmetric SO(10)}}
\end{center}
\vskip .1in
\begin{center}
Alick L. Macpherson

{\it Department of Physics, University of Alberta}

{\it  Edmonton, Alberta, Canada T6G 2J1}

\vskip .2in

\end{center}

\begin{abstract}

Evaluation of nucleon decay modes and branching ratios in a non-minimal
supersymmetric SO(10) grand unified theory is presented. The non-minimal
GUT considered is the supersymmetrised version of the `realistic' SO(10)
model originally proposed by Harvey, Reiss, and Ramond, which is realistic
in that it gives acceptable charged fermion and neutrino masses within the
context of a phenomenological fit to the low energy standard model inputs.
Despite a complicated Higgs sector, the SO(10) $\underline{10}$ Higgs
superfield mass insertion is found to be the sole contribution to the tree
level F-term governing nucleon decay. The resulting dimension 5 operators
that mediate nucleon decay give branching ratio predictions parameterised
by a single parameter, the ratio of the Yukawa couplings of the
$\underline{10}$ to the fermion generations. For parameter values
corresponding to a lack of dominance of the third family self coupling,
the dominant nucleon decay modes are $p \rightarrow K^{+} +
\bar{\nu}_{\mu}$ and $n \rightarrow K^{0} + \bar{\nu}_{\mu}$, as expected.
Further, the charged muon decay modes are enhanced by two orders of
magnitude over the standard minimal SUSY SU(5) predictions, thus
predicting a distinct spectrum of `visible' modes. These charged muon
decay modes, along with $p \rightarrow \pi^{+} + \bar{\nu}_{\mu}$ and $n
\rightarrow \pi^{0} + \bar{\nu}_{\mu}$, which are moderately enhanced over
the SUSY SU(5) prediction, suggest a distinguishing fingerprint of this
particular GUT model, and if nucleon decay is observed at Super-KAMIOKANDE
the predicted branching ratio spectrum can be used to determine the
validity of this `realistic' SO(10) SUSY GUT model. 

\end{abstract}
\baselineskip=18pt
\end{titlepage}

\baselineskip=18pt

\section{Introduction}

Nucleon decay is by definition a baryon number violating process, and
within the context of the standard model (SM) of particle physics is
forbidden \footnote{Baryon number is not conserved in the SM, as violation
occurs in weak interactions via instanton effects and the triangle
anomaly, but the rate is suppressed and also involves violation of 3 units
of baryon number for 3 standard model generations, thus making it
irrelevant to nucleon decay \cite{weakanom}.}.  Yet there is a strong
motivation for assuming that baryon number violation occurs; particularly
the fact that there is no baryonic analog of the electromagnetic gauge
invariance \cite{Lee}(which guarantees the conservation of electric
charge), the presence of a baryonic asymmetry in the Universe
\cite{Sakharov}, and the violation of baryon number conservation by black
holes \cite{BH}. Allowing baryon number violation then suggests that the
SM is only a low energy effective theory, and as such the stability of the
nucleon is brought into question. This view was further reinforced when
the adoption of Grand Unified Theories (GUTs) as a SM extension appeared
to explain a large number of questions left unanswered by the SM
\cite{GUTsolve}. The GUT scheme was introduced to attempt to unify the SM
interactions under a single simple gauge group. Imposition of such an
underlying GUT structure then provided a new mechanism by which baryon
number violation could occur, and nucleon decay induced
\cite{Georgi,Pati}. 

This nucleon decay mechanism is due to the fact that for conventional
GUTs, quarks and leptons are placed in the same multiplets of the GUT
gauge group. The coupling of these multiplets to either gauge or Higgs
boson representations then gives interactions that couple quarks to
leptons, and below the GUT scale, produce effective operators that induce
nucleon decay. These tree level operators are four fermion dimension 6
operators \cite{W79,Zee} built from two fermion-fermion-boson vertices by
means of a gauge or Higgs boson exchange. As the low energy limit of the
internal boson propagator is ${1 \over M_{G}^{2}}$ ($M_{G}$ is the mass
scale at which the GUT is spontaneously broken), the four fermion
interaction reduces at low energy to an effective four fermion vertex
scaled by two inverse powers of $M_{G}$. It is this class of effective
vertex that would mediate nucleon decay in non-supersymmetric GUT models. 

In the archetypal GUT - minimal non-supersymmetric SU(5) - first proposed
by Georgi and Glashow \cite{Georgi}, the unification scale is $M_{G} \sim
5 \times 10^{14}$ GeV \cite{GQW,Goldhaber}, and predicts the most dominant
decay mode to be $p \rightarrow \pi^{0} + e^{+}$ with a partial lifetime
of $\tau_{p} \sim 4.5 \times 10^{29 \pm 1.7}$ yrs \cite{Goldhaber}. This
is to be contrasted with the experimental lower bound obtained from IMB-3
Collaboration, of $\tau_{p} > 5.5 \times 10^{32}$ yrs
\cite{IMB,Barloutaud}. Clearly the minimal SU(5) model predicts proton
decay at too rapid a rate, thereby ruling it out as a realistic GUT
candidate. Nucleon decay channels and partial lifetime predictions have
been calculated for a variety of GUT models \cite{Ross}, including
non-minimal SU(5) (which includes a $\underline{45}$ Higgs rep in an
attempt to predict the fermion masses), minimal and non-minimal SO(10),
and an E$_{6}$ GUT model. Unfortunately, all these models tend to fail on
the basis of a unification scale $M_{G} \sim (2 - 7) \times 10^{14}$ GeV,
which implies an overly rapid nucleon decay rate as well as a prediction
for $\sin^{2} \theta_{W}$ that is inconsistent with the high precision LEP
measurements \cite{LEP}. 

As conventional GUTs are essentially condemned by these failings,
attention has turned to the supersymmetric GUT models (SUSY GUTs).
Imposing supersymmetry - a symmetry that relates bosons and fermions - has
the effect of doubling the particle content below the GUT scale, which
results in the slowing of the SM gauge coupling running, and
consequentially predicts a consistent gauge coupling unification at a
higher scale. Thus, a SUSY GUT model not only addresses the matter of the
consistency of the $\sin^{2} \theta_{W}$ prediction, but it also predicts
a unification scale that is typically two orders of magnitude larger than
that of conventional GUTs. This increase in the unification scale induces
a suppression factor of order $10^{-8}$ in the decay rates of four fermion
dimension 6 operators generated by boson exchange, placing the dimension 6
mediated nucleon partial lifetime predictions well beyond the experimental
lower bound. However, with the advent of Super-KAMIOKANDE, even the decay
mediated by the dimension 6 operators may be observable. 

Yet the extension to a SUSY GUT model permits a new operator, capable of
being the dominant contribution to nucleon decay. This operator is a
dimension 5 fermion-fermion-sfermion-sfermion effective operator
\cite{SUSYdim5,Weinberg} constructed from either two
fermion-sfermion-Higgsino vertices or a fermion-fermion-Higgs and a
sfermion-sfermion-Higgs vertex by means of a heavy colour triplet Higgino
or Higgs exchange below the GUT scale. Such an operator then evolves down
to the SUSY breaking scale, at which point the sfermions are `dressed' by
gaugino exchange to give an effective four fermion vertex that mediates
nucleon decay. As the low energy limit of the dimension 5 operator is
scaled by ${1 \over M_{G}}$, nucleon decay via this operator generally
dominates over those mediated by the conventional dimension 6
operators\footnote{It is assumed that R-parity is invoked so to rule out
dangerous dimension 4 operators.}. 

Investigations of nucleon decay in a number of SUSY GUT models have been
carried out, beginning with the supersymmetrised version of minimal SU(5)
\cite{GeorgiDimop,Sakai}. Unlike its non-SUSY cousin, this model predicts
the dominant nucleon decay modes to be $p \rightarrow K^{+} +
\bar{\nu}_{\mu}$ and $n \rightarrow K^{0} + \bar{\nu}_{\mu}$, and as the
unification scale is $M_{G} \sim 2.5 \times 10^{16}$ GeV the partial
lifetime prediction is $\tau_{p \rightarrow K^{+} + \bar{\nu}} \sim 10^{29
\pm 4}$ yrs
\cite{DRW,ENR,LL1,LL2,LL3,LL4,LL5,LL6,LL7,LL8,LL9,LL10,LL11,LL12,LL13,LL14,
LL15,LL16,LL17}
(with $M_{H_{3}}$ set to $M_{G}$). This prediction does not disagree with
the experimental bound of $\tau_{p \rightarrow K^{+} + \bar{\nu}} >
10^{32}$ yrs (obtained from the water \v{C}erenkov detector of the
KAMIOKANDE Collaboration \cite{Kamiokande,Barloutaud}) due mainly to the
large uncertainty resulting from the value of the Higgs/Higgsino colour
triplet mass. Likewise, predictions for non-minimal SUSY SU(5), minimal
SUSY SO(10) result in a marginal degree of compatibility with the
experimental lower bounds on the partial lifetimes of the various nucleon
decay channels \cite{Ross}. This marginal consistency suggests that an
improvement on the experimental lower bounds could lead to either a
rejection of nucleon decay via SUSY GUT generated dimension 5 operators,
or an observation of nucleon decay. Yet the uncertainties in nucleon
partial lifetime predictions preclude model discrimination by rate. In
order to distinguish the underlying SUSY GUT structure, the relative decay
rate predictions within a model should be determined, and then used to
identify the SUSY GUT candidate, once nucleon decay has been observed. 

With this strategy in mind, this paper presents the branching ratios for
nucleon decay in a particular `realistic' SUSY GUT model. The model chosen
is a supersymmetrised version \cite{BDO} of the non-minimal SO(10) GUT
proposed by Harvey, Reiss, and Ramond \cite{HRR}, which was constructed
primarily to reproduce a consistent phenomenological fit to the observed
SM fermion masses and mixing angles. This realistic non-minimal SUSY
SO(10) model, like its non-SUSY counterpart, can be viewed as a
sophisticated phenomenological one, as it supports a rather expansive
Higgs sector that is responsible for the required Yukawa coupling texture.
It will be shown that analysis of the various nucleon decay channels
mediated by the dimension 5 operators of this model results in branching
ratio predictions depending on a single parameter, with the branching
ratios for some observable modes enhanced by factors of order 100 over the
minimal SUSY SU(5) predictions. This in turn suggests that if nucleon
decay is observed at Super-KAMIOKANDE, the $p \rightarrow K^{0} +
\mu^{+}$, $p \rightarrow \pi^{0} + \mu^{+}$, and $n \rightarrow \pi^{-} +
\mu^{+}$ decay channel may play a significant role in identifying the
structure of the underlying SUSY GUT.

In this paper, section 2 presents the non-minimal SUSY SO(10) model to be
used, section 3 examines in detail the low energy quark-level effective
lagrangian, while section 4 discusses the effective lagrangian at the
hadronic level and presents the branching ratio predictions. Finally, in
section 5 a discussion of these predictions and the conclusions that can
be drawn from them is given. 

\section{The Non-Minimal SUSY SO(10) Model} 

As mentioned, this analysis is based on the non-minimal SO(10) GUT model
of Harvey, Reiss, and Ramond \cite{HRR}, which has been explicitly
constructed to generate a mass spectrum (including mixing angles) of the
SM fermions from the GUT. An advantage of the choice of SO(10) as the
gauge group is that the lowest dimensional chiral representation that
accommodates the observed SM fermions is the $\underline{16}$, which
allows for the assignment of one family of SM fermions plus a right handed
neutrino, and does not include any mirror fermions. This in turn places
constraints on the possible Higgs sector representations, since the
fermion masses transform under SO(10) as $\underline{16} \times
\underline{16} = (\underline{10} + \underline{126})_{S} +
\underline{120}_{A}$, (where $ S$ and $ A$ refer to the symmetric and
antisymmetric parts respectively), implying that the allowed Higgs reps
that couple to fermions to form SO(10) invariant Yukawa terms are the
$\underline{10}, \underline{120},$ and the $\overline{\underline{126}}$.
The `realistic' model of Harvey et al. \cite{HRR} is then constructed from
the representations in such a way that SO(10) GUT is broken directly to
the SM gauge structure of $SU(3)_{C} \times SU(2) \times U(1)$, and the
GUT scale texture of Yukawa couplings incorporates the up quark mass
matrix ansatze of Fritzsch \cite{Fritzsch} and the down quark and charged
lepton mass matrix ansatze of Georgi and Jarlskog \cite{Georgi-Jarlskog}
in such a way that the Oakes relation \cite{Oakes} results. The cost of
such a model is the expansion of the Higgs sector well beyond that of most
minimal models. The particle content of this model is given in terms of a
$\underline{45}$ that is the adjoint of vector bosons, three families of
fermions ($\underline{16}_{1}, \underline{16}_{2},\underline{16}_{3}$),
and a scalar sector composed of a $\underline{54}$, a complex
$\underline{10}$, and three families of $\underline{126}$
($\underline{126}_{1}, \underline{126}_{2},\underline{126}_{3}$) - all of
which are required for a viable spectrum of fermion masses. Note that the
phenomenologically observed mass spectrum can be produced without
requiring the presence of the $\underline{120}$ rep, which has a Yukawa
coupling to the fermions that is antisymmetric in generation indices (as
the SM fermions are expressed in terms of a single chirality, and the spin
0 fields occur in a product that is symmetric in Lorentz indices). 

The extension \cite{BDO} of this model to that of a SUSY SO(10) model
is straight forward, as the gauge, fermion and conjugate Higgs fields are
converted into vector and chiral superfields, giving a superfield content
of: 

\begin{table}[h]
\begin{center}
\begin{tabular}{rcl}
Vector superfields&:& $\underline{45}$\\ 
Chiral superfields&: & $\underline{10},\underline{16}_{1},
\underline{16}_{2},\underline{16}_{3},\underline{54},
\overline{\underline{126}}_{1},\overline{\underline{126}}_{2}, 
\overline{\underline{126}}_{3}$
\end{tabular}
\end{center}
\end{table}
This SUSY SO(10) model, like its non-SUSY counterpart, is distinguished by
its sophisticated Yukawa texture, composed of the $\underline{10},
\underline{16}$, and $\overline{\underline{126}}$ chiral superfield reps. 
Specifically, the model is defined in terms of its superpotential, and for
the purposes of nucleon decay, the relevant terms of the superpotential for
this SO(10) model are
\begin{eqnarray} 
W &=& (A \underline{16}_{1} \times \underline{16}_{2} + B
\underline{16}_{3} \times \underline{16}_{3}) \times
\overline{\underline{126}}_{1} + (a \underline{16}_{1} \times
\underline{16}_{2} + b \underline{16}_{3} \times \underline{16}_{3})
\times \underline{10} \nonumber \\ 
&& + c (\underline{16}_{2} \times
\underline{16}_{2}) \times \overline{\underline{126}}_{2} + d
(\underline{16}_{2}
\times \underline{16}_{3}) \times \overline{\underline{126}}_{3} + M_{G}
\underline{10} \times \underline{10}  
\label{WSO10}
\end{eqnarray} 
with the superpotential expressed in terms of the SO(10) representations,
and $A,B,a,b,c,d$ as the undetermined GUT scale Yukawa couplings. 

\begin{figure}[ht]
\includegraphics{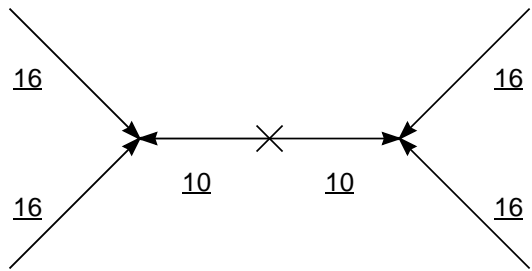}
\vspace{3.4cm}

\caption[]{The only F-term supergraph that contributes to nucleon decay.}

\label{supergraph}
\end{figure}

The beauty of this globally supersymmetric model is that as
$\underline{10} \times \underline{10} \supset \underline{1}$ and
$\overline{\underline{126}} \times \overline{\underline{126}} \supset
\hspace{-0.45cm} \slash \hspace{+0.2cm} \underline{1}$, the only SO(10)
invariant F-term that contributes to nucleon decay below the spontaneously
broken SO(10) GUT scale is given by Figure \ref{supergraph}. 

The key point here is that this superfield diagram has a
Higgs/Higgsino mass insertion that involves only the $\underline{10}$ (the
$\underline{120}$ reps are absent!), which implies that only the GUT scale
Yukawa couplings of the $\underline{16}$'s to the $\underline{10}$ are of
relevance to the predictions of nucleon decay (i.e. $ a$ and $ b$ in
equation (\ref{WSO10}) are the only relevant couplings). In terms of the
particle diagrams, the only tree level diagrams of concern are

\begin{figure}[hb]
\includegraphics{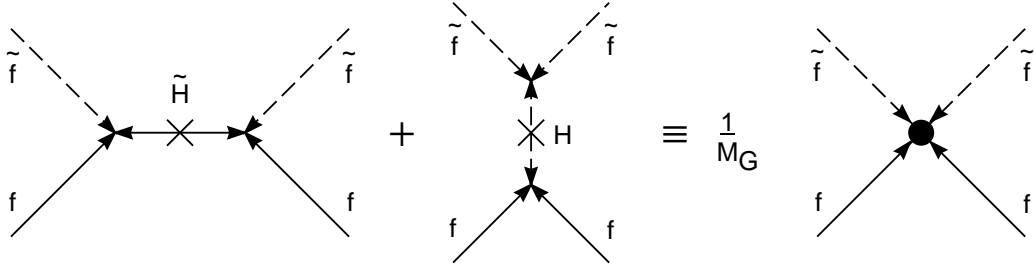}
\vspace{3.2cm}

\caption[]{The two particle diagrams generated by the F-term supergraph of
Figure \ref{supergraph}.}

\label{trees}
\end{figure}
Here the first dimension 5 diagram exhibits the exchange of a Higgsino of
GUT scale mass ($M_{G}$), and so for momentum below the GUT scale the
Higgsino propagator reduces to a factor of ${1\over M_{G}}$. The second
diagram in Figure \ref{trees} involves the exchange of a GUT scale Higgs
scalar whose propagator reduces to ${1\over M_{G}^{2}}$, but due to the
weighting of the sfermion-sfermion-Higgs trilinear coupling by one power
of $M_{G}$, the resulting diagram also contributes to the dimension 5
operator with weight ${1 \over M_{G}}$. Thus the effective dimension 5
operator, valid between the SUSY GUT scale and the SUSY breaking scale
(assumed to be of order the electroweak scale) is a combination of both
diagrams, and is represented by the effective vertex in Figure
\ref{trees}. 

\section{The Dimension 5 Operators}

In order to evaluate these dimension 5 operator contributions to nucleon
decay, the superpotential must be re-expressed in terms of the superfields
corresponding to the SM content. This may be done in a two step process,
which first involves the re-expression of the superpotential in a compact
SU(5) notation, followed by a decomposition of the SU(5) superfields into
their SM components. Such a decomposition can be used as the F-term of
Figure \ref{supergraph} is the only dimension 5 contribution to nucleon
decay, and it relies only on the Higgs $\underline{10}$ of SO(10) which
has the SU(5) decomposition $\underline{10} \rightarrow \underline{5} +
\overline{\underline{5}}$. (Note that the SU(5) decomposition of the
$\underline{16}$ is $\underline{16} \rightarrow \underline{10} +
\overline{\underline{5}} + \underline{1}$.) Thus the superpotential terms
that contribute to nucleon decay can be written as
\begin{equation}
W_{SU(5)}=\sqrt{2} \chi_{a}^{\alpha \beta} M_{ab}^{D} \psi_{b \alpha} H_{2
\beta} - {1 \over 4} \epsilon_{\alpha \beta \gamma \delta \epsilon}
\chi^{\alpha \beta}_{c} M^{U}_{cd} \chi^{\gamma \delta}_{d}
H_{1}^{\epsilon} + M_{G} H_{1}^{\alpha} H_{2 \alpha}
\label{WSU5} 
\end{equation}
with $ W_{SU(5)}$ being valid at the GUT scale. Here $M^{U}$ and $M^{D}$
are $3 \times 3$ matrices in generation space that express in a compact
form the Yukawa coupling texture expressed in equation (\ref{WSO10}).
Below the SO(10) scale, the heavy Higgs superfield can be integrated out
to give an effective superpotential (that is appropriate below the GUT
scale but above the SUSY breaking scale).  This effective superpotential
is
\begin{equation}
W^{\mbox{eff}}_{SU(5)} = {\sqrt{2} \over 4 M_{G}} \epsilon_{\alpha
\beta  \gamma \delta i} \chi^{\alpha \beta}_{a} M^{U}_{ab} \chi^{\gamma
\delta}_{b} \chi^{i \epsilon}_{c} M^{D}_{cd} \psi_{d \epsilon}
\label{WeffSU5}
\end{equation}

Here the Greek indices $\alpha, \beta,$ ... are SU(5) indices, the family
indices are $(a,b,c,d)$, and the index $i$ runs from 1 to 3. Also, the
Lorentz structure is suppressed, so to focus on the generation and SU(5)
structure. Restriction to the tree level diagrams relevant to nucleon decay
(Figure \ref{supergraph}) then implies the
Yukawa texture matrices for this effective superpotential are of the form
\begin{equation}
M^{U} = \left [ \matrix{
0 & a & 0 \cr
a & 0 & 0 \cr
0 & 0 & b \cr } \right ] = M^{D}
\label{Yuk}
\end{equation}

As it is assumed that this SUSY SO(10) breaks straight to the minimally
supersymmetric standard model (MSSM), the superpotential can then be
further decomposed into the SM quark and lepton superfields. Typically the
decomposition for one family of left-handed SU(5) matter superfields are
\begin{eqnarray}
\psi_{\alpha}= \left [ \matrix{
D_{1}\cr
D_{2}\cr
D_{3}\cr
l\cr
- \nu\cr
}\right ]_{L}
& &\chi^{\alpha \beta} = {1 \over \sqrt{2}} 
\left [\matrix{
0 & U_{3}& -U_{2} & -u^{1} & -d^{1}\cr
-U_{3}& 0 & U_{1} & -u^{2} & -d^{2}\cr
U_{2} & -U_{1}& 0 & -u^{3} & -d^{3}\cr
u^{1} & u^{2}& u^{3} & 0 & -L\cr
d^{1} & d^{2}& d^{3} & L & 0\cr
}\right ]_{L} 
\label{qlcontent}
\end{eqnarray}
where $ U_{i}, D_{i}$ and $ L_{i}$ are the charge conjugations of the
right handed SU(2) singlet up, down, and charged lepton fields.
Substitution of this decomposition into the superpotential (equation
(\ref{WeffSU5})), results in an effective superpotential relevant to nucleon
decay that is expressed in chiral superfields associated with the SM. The
form of the effective superpotential in question is
\begin{eqnarray}
W_{\mbox{eff}}^{\mbox{SM}} &=& {-1 \over 2 M_{G}} [ {\epsilon^{ijk}
\over 4} L_{a} M^{U}_{ab} U_{bi} U_{ck} M^{D}_{cd} D_{dj} + {\epsilon^{ijk}
\over 4} U_{ai} M^{U}_{ab} L_{b} U_{ck} M^{D}_{cd} D_{dj} \\ \nonumber 
&&- \epsilon_{ijk} (u_{a}^{i} M^{U}_{ab} d_{b}^{j} - d_{a}^{i} M^{U}_{ab}
u_{b}^{j}) (u_{c}^{k} M^{D}_{cd} l_{d} - d_{c}^{k} M^{D}_{cd} \mu_{d}) ] 
\label{WeffSM}
\end{eqnarray}

\begin{figure}[hbt]
\includegraphics{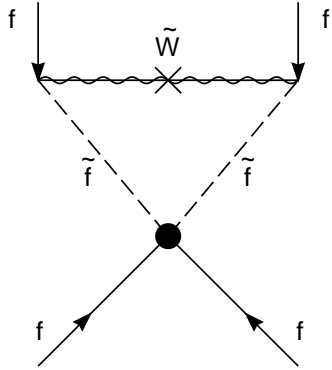}
\vspace{4.7cm}

\caption[]{The particle diagram for the nucleon decay operators after being
dressed by the wino or charged higgsino exchange.}

\label{dressdiag}
\end{figure}

{}From this superpotential it is clear that as of a result of the SU(2)
content, there are two classes of F-terms; the $(\mbox{LLLL})_{F}$ and the
$(\mbox{RRRR})_{F}$ terms (here the notation of reference \cite{Weinberg}
is used to emphasise the SU(2) weak content of the operators). However as
the $(\mbox{RRRR})_{F}$ terms are antisymmetric in generation indices
($a,b,c$) - due to the Bose statistics of superfields in a superpotential
- their composition is such that they must contain either a charm or a top
SU(2) singlet superfield. This superfield generation remains, to a first
approximation, unchanged on the dressing of the operator by gluino or bino
exchange at the SUSY breaking scale (SU(2) gaugino exchange is forbidden
for these singlet superfields), and so the low energy four fermion
operator contains either a charm or a top quark. This implies that the
($\mbox{RRRR})_{F}$ term contribution to nucleon decay is suppressed,
leaving only the $(\mbox{LLLL})_{F}$ terms. The effective lagrangian
relevant to nucleon decay is then obtained from the $(\mbox{LLLL})_{F}$
term of the superpotential by the usual method (${\cal L}_{\mbox{Int}} =
{1 \over 2} \sum_{i,j} ({\partial W \over \partial \Phi_{i} \partial
\Phi_{j}} |_{\Phi=\phi} \psi_{i} \psi_{j} + h.c.) - \sum_{i} |{\partial W
\over \partial \Phi_{i} }|^{2}_{\Phi = \phi}$ with $\Phi$ representing a
chiral superfield, and $\phi$ and $\psi$ the scalar and fermionic parts),
which results in vertices composed of two particles and two sparticles.
These vertices are then renormalisation group evolved down to the SUSY
breaking scale ($\sim O(M_{W})$) at which point the dimension 5 operator
is converted to a dimension 6 operator via gaugino and Higgsino exchanges.
This dressing is schematically shown in Figure \ref{dressdiag}. 
 
Of all the gaugino and higgsino exchanges associated with the dressing of
the $(\mbox{LLLL})_{F}$ dimension 5 operators, the dominant contribution
comes from the charged wino. The gluino, neutral gaugino, and neutral
Higgsino exchange contributions to the dressed operator are suppressed as
their exchange is approximately generation diagonal and their contribution
is thereby suppressed due to small Yukawa couplings of the first and
second generation fields present in the dimension 5 operators
\cite{DRW,ENR}. The charged-Higgsino exchanges are also suppressed, due to
their Higgs strength Yukawa couplings to the first and and second
generation fermions, thereby leaving the charged wino as the dominant
contribution to the loop integral. Complete calculations of the loop
dressing by chargino eigenstates have been performed by Sakai
\cite{Sakai}, and their implications for nucleon decay rates have been
explored for minimal and non-minimal versions of SU(5) \cite{LL8}. Here
however, the simplifying assumption of wino dominance of the decay
amplitude is invoked. Preforming the loop integration, results in a
triangle diagram factor, that although it depends on mass eigen-values and
mixing angles of the sparticles in the loop, can be approximated (in the
pure charged wino exchange limit) by \cite{Sakai}
\begin{eqnarray}
{\alpha_{2} \over 2 \pi} f(\tilde{u},\tilde{d},\tilde{W}) &=& g_{2}^{2}
\int {d^{4}k \over i (2 \pi)^{4}} {1 \over m_{\tilde{u}}^{2} - k^{2}} {1
\over m_{\tilde{d}}^{2} - k^{2}} {1 \over m_{\tilde{W}} - \slash
\hspace{-0.22cm} k} \\ \nonumber &\simeq& {\alpha_{2} \over 2 \pi}
{m_{\tilde{W}} \over m_{\tilde{u}}^{2} -m_{\tilde{d}}^{2}} \left (
{m_{\tilde{u}}^{2} \over m_{\tilde{u}}^{2} - m_{\tilde{W}}^{2}} \ln
{m_{\tilde{u}}^{2} \over m_{\tilde{W}}^{2}} -{m_{\tilde{d}}^{2} \over
m_{\tilde{d}}^{2} -m_{\tilde{W}}^{2}} \ln {m_{\tilde{d}}^{2} \over
m_{\tilde{W}}^{2}} \right )
\label{triangle}
\end{eqnarray}
and so becomes a multiplive factor of the dressed four fermion
operator. 

{}From the superpotential (equation (\ref{WeffSM})) the effective lagrangian
for the dressed quark level operators of Figure (\ref{dressdiag}) can be
obtained, and with the use of equation (\ref{triangle}) it has the form
\begin{eqnarray} 
{\cal L} &=&{\alpha_{2} \over 2 \pi M_{G}} R_{S} R_{L} M^{U}_{ab}
M^{D}_{cd} \epsilon_{ijk} \left[ (u^{i}_{a} d_{b}^{j}) (d^{k}_{c} \nu_{d})
\{f(u_{c},l_{d},m_{\tilde{W}}) + f(u_{a}, d_{b}, m_{\tilde{W}})\} \right. 
\nonumber \\ &&+ (d^{i}_{a} u_{b}^{j}) (u^{k}_{c} l_{d})
\{f(d_{c},\nu_{d},m_{\tilde{W}}) + f(d_{a}, u_{b},
m_{\tilde{W}})\} \nonumber \\ &&+
(u^{i}_{b} d_{c}^{j}) (u^{k}_{a} l_{d}) \{f(u_{c},d_{b},m_{\tilde{W}}) +
f(d_{a}, \nu_{d}, m_{\tilde{W}})\} \left. \right. \\ &&+ (d^{i}_{a}
\nu_{d}) (d^{j}_{b} u_{c}^{k}) \{f(u_{a},l_{d},m_{\tilde{W}}) + f(u_{b},
d_{c}, m_{\tilde{W}})\} \left. \right] + h.c.  \nonumber
\label{dressedL} 
\end{eqnarray}
Here the $R_{S}$ and $R_{L}$ are the short and long range renormalisation
factors. The short range renormalisation accounts for the renormalisation
effects from the SO(10) to the SUSY breaking scale, while the long range
factor is from the SUSY braking scale to a low energy scale (assumed here
to be 1 GeV). $R_{S}$ can be shown to be generation independent, and
can be taken to be \cite{DRW,ENR} 
\begin{eqnarray}
R_{S} &=& 
\left[ {\alpha_{3}(m_{S}) \over \alpha_{G}}\right]^{-4 \over 9}
\left[ {\alpha_{2}(m_{S}) \over \alpha_{G}}\right]^{-3 \over 2}
\left[ {\alpha_{1}(m_{S}) \over \alpha_{G}}\right]^{5 \over 396} \simeq 0.91
\label{RS}
\end{eqnarray} 
where $m_{S}$ is the SUSY breaking scale (which here, has been set to the
electroweak scale $m_{W}$). The long range renormalisation is
predominantly a result of QCD interactions between the SUSY scale and 1
GeV, and encompasses the renormalisation of the Yukawa couplings and
anomalous dimension corrections to the four fermion operators. Again,
following reference \cite{ENR},
\begin{eqnarray}
R_{L} &\simeq& 
\left[ {\alpha_{3}(1GeV) \over \alpha_{3}(m_{c})}\right]^{-2 \over 3}
\left[ {\alpha_{3}(m_{c}) \over \alpha_{3}(m_{b})}\right]^{-18 \over 25}
\left[ {\alpha_{3}(m_{b}) \over \alpha_{3}(m_{Z})}\right]^{-18 \over
23} \simeq 0.22 \label{RL}
\end{eqnarray}

This effective lagrangian, as written, is for four fermion operators with
the quarks and leptons expressed in their gauge interaction
eigenstates; this however is easily remedied by
rotating from a gauge interaction to a mass eigenstate basis.  Due to the
mismatch in the rotations of the charge ${2 \over 3}, -{1 \over 3}$, and
$-1$ fields, the operators incur additional generation mixing. The
rotation matrices appropriate to this model are obtained from the
diagonalisation of the mass matrices which are defined in terms of the
Yukawa coupling texture (as specified in equation (\ref{WSO10})) and the
vevs of the various Higgs reps. For these $\Delta I_{W} = {1 \over 2}$
Dirac masses ($I_{W}$ denotes weak isospin), it is convenient to consider the
SO(10)
vev contributions in terms of their SU(5) content. The contribution of
SU(5) vevs to the quark and charged lepton masses is as follows \cite{HRR}: 
\begin{eqnarray}
< \dots > &\sim& \underline{5} \mbox{ gives a contribution to the charge 
${2 \over 3}$ mass} \nonumber \\
< \dots > &\sim& \overline{\underline{5}} \mbox{ gives an
equal weight contribution to the charge $-{1 \over 3}$ and  $-1$ masses}
\nonumber \\
< \dots > &\sim& \overline{\underline{45}} \mbox{ gives a contribution of
relative weight $1:-3$  to the charge $-{1 \over 3}$ and  $-1$ masses}
\nonumber 
\label{vevs}
\end{eqnarray}
The SO(10) Higgs vevs structure is then decomposed as
\begin{eqnarray}
< \underline{10} > &=& r (\mbox{along } \overline{\underline{5}}) + p
(\mbox{along } \underline{5})  \nonumber \\
< \overline{\underline{126}_{1}} > &=& t (\mbox{along } \underline{5}) 
\nonumber \\
< \overline{\underline{126}_{2}} > &=& s (\mbox{along }
\overline{\underline{45}}) \\
< \overline{\underline{126}_{3}} > &=& q (\mbox{along } \underline{5}) 
\nonumber
\label{Higgsvev}
\end{eqnarray}
where $p,q,r,s,t$ are taken as complex vevs. The assumption of complex
vevs allows for the generation of soft CP violation through the process of
symmetry breaking. Yet it is assumed that soft CP
violation is not the sole source
of CP violation in the model. Hard CP violation is also permitted due to the
fact that, unlike the non-SUSY model of reference \cite{HRR}, the Yukawa
couplings of equation (\ref{WSO10}) are taken to be complex \cite{BDO}. 
 
As the masses in the low energy effective SUSY theory arise from the
Yukawa couplings of the quarks and charged leptons to a single light Higgs
doublet of $\Delta I_{W} = {1 \over 2}$, the mass matrices can be
formulated in terms of the vev of this light Higgs. From the SU(5)
decomposition of the SO(10) Higgs vevs, this light Higgs vev is a linear
combination of the doublets in the $\underline{10},
\overline{\underline{126}}_{1}, \overline{\underline{126}}_{2}$, and
$\overline{\underline{126}}_{3}$ (in the ratio $|r + p| : t : s : q$), and
so the GUT scale couplings appearing in the mass matrices can be read off. 
Yet as it is the mass texture at the SUSY breaking scale that must be
diagonalised, the entries in these Yukawa coupling texture matrices at the
GUT scale must be evolved down to the SUSY scale via the renormalisation
group equations, as done in Dimopoulos, Hall, and Raby \cite{DHR} for
`realistic' Yukawa matrices of this form.  The
quark and charged lepton Yukawa matrices specified at the SUSY breaking
scale are then the mass matrices that are diagonalised. From equations
(\ref{WSO10}) and (\ref{Higgsvev}) the GUT and SUSY scale mass matrix textures
of the quarks and charged leptons are:  
\begin{equation}
\begin{array}{ccc}
\mbox{GUT scale texture} & & \mbox{SUSY scale texture} \\ \nonumber
U =\left [ \matrix{
0 & P_{G} & 0 \cr
P_{G} & 0 & Q_{G} \cr
0 & Q_{G} & V_{G} \cr
}\right ] 
& \longrightarrow &
U =\left [ \matrix{
0 & P & 0 \cr
P & \delta _{u} & Q \cr
0 & Q & V \cr
}\right ] 
\\ 
D=\left [\matrix{
0 & R_{G} e^{i \varphi_{G}} & 0 \cr
R_{G} e^{-i \varphi_{G}} & S_{G} & 0 \cr
0 & 0 & T_{G} \cr
}\right ]
& \longrightarrow &
D=\left [\matrix{
0 & Re^{i \varphi} & 0 \cr
Re^{-i \varphi} & S & \delta_{d} \cr
0 & 0 & T \cr
}\right ]
\\ 
L = \left [\matrix{
0 & R_{G} & 0 \cr
R_{G} & -3 S_{G} & 0 \cr
0 & 0 & T_{G} \cr
}\right ]
& \longrightarrow&
L = \left [\matrix{
0 & R & 0 \cr
R & -3 S & 0 \cr
0 & 0 & T \cr
}\right ]
\end{array}
\label{masstexture}
\end{equation}
with the assignments
\begin{eqnarray}
&\begin{array}{cc}
P = a p + A t  &V = b p + B t
\end{array}& \nonumber \\
&\begin{array}{cccc}
R = a r & T = b r & S = c s & Q= d q 
\end{array}&
\label{definitions}
\end{eqnarray}
and the subscript $G$ indicating entries defined at the GUT scale. Here,
the zero entries in the mass textures are the result of accidental
discrete symmetries, which if broken, allow the generation of non-zero
entries by means of the renormalisation group equations as the mass
matrices are renormalised down to lower energies. This is indeed the case
for the entries $\delta_{u}$ and $\delta_{d}$ which occur due to the
violation of a discrete symmetry at the GUT scale. 

Although both the Yukawa couplings and the SO(10) Higgs vevs are complex,
thereby permitting both hard and soft CP violation, the entries in the
mass matrix textures, as given in (\ref{masstexture}), have been
rendered explicitly real by means of quark and charged lepton field
redefinitions. It is then these (SUSY scale) matrices, with 8 real
parameters and one phase, that are diagonalised and the mass eigenvalues
fitted to the low energy data, following  Dimopoulos, Hall, and Raby
\cite{DHR}. The diagonalisation proceeds by means of unitary and biunitary
transformations of the form $U^{\mbox{diag}}= V_{u} U V_{u}^{\dagger}$,
$D^{\mbox{diag}}= V^{L}_{d} D V_{d}^{R \dagger}$, and $L^{\mbox{diag}}=
V_{l} L V_{l}^{\dagger}$, and in following the assumptions of reference
\cite{DHR}, that $ V >> Q \sim \delta_{u} >> P $ and $ T >> S \sim
\delta_{d} >> R$, the approximate mixing matrices are of the form
\begin{eqnarray}
V_{u} &=&
\left [ \matrix{
c_{2} & s_{2} & 0 \cr
-s_{2} & c_{2} & 0 \cr
0 & 0 & 1 \cr
}\right ] 
\left [ \matrix{
1 & 0 & 0 \cr
c_{3} & s_{3} & 0 \cr
-s_{3} & c_{3} & 0 \cr
}\right ]
\nonumber \\
V^{L}_{d} &=&
\left [ \matrix{
c_{1} & -s_{1} & 0 \cr
s_{1} & c_{1} & 0 \cr
0 & 0 & 1 \cr
}\right ] 
\left [ \matrix{
1 & 0 & 0 \cr
c_{4} & s_{4} & 0 \cr
-s_{4} & c_{4} & 0 \cr
}\right ]
\left [ \matrix{
1 & 0 & 0 \cr
0 & e^{i \varphi} & 0 \cr
0 & 0 & e^{i \varphi} \cr
}\right ] 
\\
V_{l} &=&
\left [ \matrix{
c_{5} & s_{5} & 0 \cr
-s_{5} & c_{5} & 0 \cr
0 & 0 & 1 \cr
}\right ] \nonumber
\label{rotations}
\end{eqnarray} 
with $s_{i} = \sin \theta_{i}$ and $c_{i} = \cos \theta_{i}$. The angles
define in these rotation matrices can then be determined by fitting the
mass eigenvalues to the low energy data. Using the low energy input data
of \cite{DHR}, the resulting phenomenological
fit specifies the angles as
\begin{eqnarray}
s_{1} \simeq 0.196 & s_{2} \simeq 0.05  & s_{3} \simeq 0.046 \\ \nonumber
s_{4} \simeq 0.0066 & s_{5} \simeq 0.070 & \cos \varphi
\simeq 0.41_{-0.15}^{+0.22}  \label{angles}
\end{eqnarray}
This phenomenological fit may need some revision in
view of the subsequent and more precise low energy data (especially in
light of the recent improvement to the bounds on the CKM matrix entry
$V_{cb}$ \cite{Vcb}) but it is expected that any revisions will have small
effects on our results, and so we continue to use the original fit. 

\section{The Hadronic Lagrangian and Branching Ratio Predictions}

With the mass eigenstate rotations defined by equations (\ref{rotations})
and (\ref{angles}), the low energy effective lagrangian of equation
(\ref{dressedL}) can be explicitly evaluated in terms of the dimension 6
four fermion quark level operators. In focusing on nucleon decay, these
quark level $(qqql)$ operators can be restricted by energy conservation,
to have a quark composition of only the $u$, $d$, and $s$ quarks. This in
turn results in only five distinct operators, namely
\begin{eqnarray}
O^{q}(dud\nu_{a})= \epsilon_{ijk} (d^{i} u^{j}) (d^{k} \nu_{a}) 
&O^{q}(sud\nu_{a})= \epsilon_{ijk} (s^{i} u^{j}) (d^{k} \nu_{a}) 
&O^{q}(uds\nu_{a})= \epsilon_{ijk} (u^{i} d^{j}) (s^{k} \nu_{a}) 
\nonumber \\
O^{q}(duul_{a})= \epsilon_{ijk} (d^{i} u^{j}) (u^{k} l_{a}) 
&O^{q}(suul_{a})= \epsilon_{ijk} (s^{i} u^{j}) (u^{k} l_{a}) 
&
\label{quarkop}
\end{eqnarray}
Thus, the effective Lagrangian, expressed at the quark level, can then be
written as ${\cal L}_{\mbox{nucleon}} =\sum C(qqql) O^{q}(qqql)$.
Here the $C(qqql)$'s are the coefficients of the distinct quark level
operators $O^{q}(qqql)$, and are determined from equation
(\ref{dressedL}) by summing the coefficients of the equivalent four fermion
effective operators, modulo Fierz transformations.  By classifying nucleon
decay in terms of its various allowed channels, the effective lagrangian
for nucleon decay can be written as
\begin{eqnarray}
{\cal L} ( n,p \rightarrow \pi + \bar{\nu}_{i})&=& C(dudv_{i})
O^{q}(dud\nu_{i}) \\ \nonumber
{\cal L} ( n,p \rightarrow \pi + l^{+}_{i})&=& C(duul_{i})
O^{q}(duul_{i}) \\ \nonumber
{\cal L} ( n,p \rightarrow  K^{0} + l^{+}_{i}) &=& C(suul_{i})
O^{q}(suul_{i}) \\ \nonumber
{\cal L} ( n,p \rightarrow K^{+} + \bar{\nu}_{i}) &=& 
C(sud\nu_{i}) O^{q}(sud\nu_{i}) + C(dus\nu_{i}) O^{q}(dus\nu_{i})
\label{Ldecay}
\end{eqnarray}

Yet these effective lagrangian contributions are in terms of quark level
operators, and so inappropriate for hadronic decay rate calculations.
Instead, they must be converted to effective lagrangian contributions at
the hadronic level, thereby permitting evaluation of the nucleon decay
rates, which although calculated at the hadronic level, are expressed in
terms of the coefficients of the quark level four fermion effective
operators specified by equations (\ref{dressedL}) and (\ref{Ldecay}).  This
conversion may be preformed using the chiral lagrangian techniques developed in
references \cite{Claudson}, \cite{LL3}, and \cite{LL4}, which express
general hadronic
level decay rates in terms of coefficients of generic four fermion quark
level operators. The results of
these decay rate calculations, in the notation of \cite{LL13}, are as
follows:  
\begin{eqnarray}
\Gamma(p \rightarrow K^{+} + \bar{\nu}_{i})&=& {(m_{p}^{2} -
m_{K}^{2})^{2} \over 32 \pi m_{p}^{3} f_{\pi}^{2}} \left |{2
m_{p} \over 3 m_{B}} {\cal D} C(sud\nu_{i}) + \left [ 1 + {m_{p} \over 3 m_{B}}
({\cal D} + 3{\cal F}) \right ] C(dus\nu_{i}) \right |^{2}
\nonumber \\
\Gamma(p \rightarrow \pi^{+} + \bar{\nu}_{i})&=& {m_{p} \over 32 \pi
f_{\pi}^{2}} \left | [1 +{\cal D} + {\cal F}] C(dud\nu_{i}) \right |^{2}
\nonumber \\
\Gamma(p \rightarrow K^{0} + l^{+}_{i})&=& {(m_{p}^{2} - m_{K}^{2})^{2}
\over 32 \pi m_{p}^{3} f_{\pi}^{2}} \left | \left [ 1 - {m_{p} \over
m_{B}} ({\cal D} - {\cal F}) \right ] C(suul_{i}) \right |^{2} 
\nonumber \\ 
\Gamma(p \rightarrow \pi^{0} + l^{+}_{i})&=& {m_{p} \over 64 \pi
f_{\pi}^{2}} \left | [ 1 + {\cal D} + {\cal F} ] C(duul_{i}) \right |^{2}
 \\
\Gamma(p \rightarrow \eta + l^{+}_{i})&=&  {3 (m_{p}^{2} - m_{\eta}^{2})^{2}
\over 64 \pi m_{p}^{3} f_{\pi}^{2}} \left | \left [ 1 - {1 \over
3} ({\cal D} - 3{\cal F}) \right ] C(duul_{i}) \right |^{2} 
\nonumber \\
\Gamma(n \rightarrow K^{0} + \bar{\nu}_{i})&=& {(m_{n}^{2} -
m_{K}^{2})^{2} \over 32 \pi m_{n}^{3} f_{\pi}^{2}} \left | \left [ 1 -
{m_{n} \over 3m_{B}}({\cal D} -3 {\cal F}) \right ] C(sud\nu_{i})
\right. \nonumber \\ &&\left. \hspace{2.05cm}+ \left [
1 + {m_{n} \over 3 m_{B}}
({\cal D} + 3{\cal F}) \right ] C(dus\nu_{i}) \right |^{2}
\nonumber \\
\Gamma(n \rightarrow \pi^{0} + \bar{\nu}_{i})&=& {m_{n} \over 64 \pi
f_{\pi}^{2}} \left | [1 +{\cal D} + {\cal F}] C(dud\nu_{i}) \right |^{2}
\nonumber \\
\Gamma(n \rightarrow \pi^{-} + l^{+}_{i})&=& {m_{n} \over 32 \pi
f_{\pi}^{2}} \left | [ 1 + {\cal D} + {\cal F} ] C(duul_{i}) \right |^{2}
\nonumber \\
\Gamma(n \rightarrow \eta + \bar{\nu}_{i})&=&  {3 (m_{n}^{2} -
m_{\eta}^{2})^{2}
\over 64 \pi m_{n}^{3} f_{\pi}^{2}} \left | \left [ 1 - {1 \over
3} ({\cal D} - 3{\cal F}) \right ] C(dud\nu_{i}) \right |^{2} 
\nonumber
\label{decayrates}
\end{eqnarray}
Here $m_{B} \equiv m_{\Sigma} = m_{\Lambda}= 1150$ MeV is the mass to be
associated with the virtual baryon exchange, $m_{n} = m_{p}$ is the
nucleon mass, and ${\cal D}= 0.81$ and ${\cal F} = 0.44$ are numerical
factors. 

\begin{figure}[t]
\includegraphics{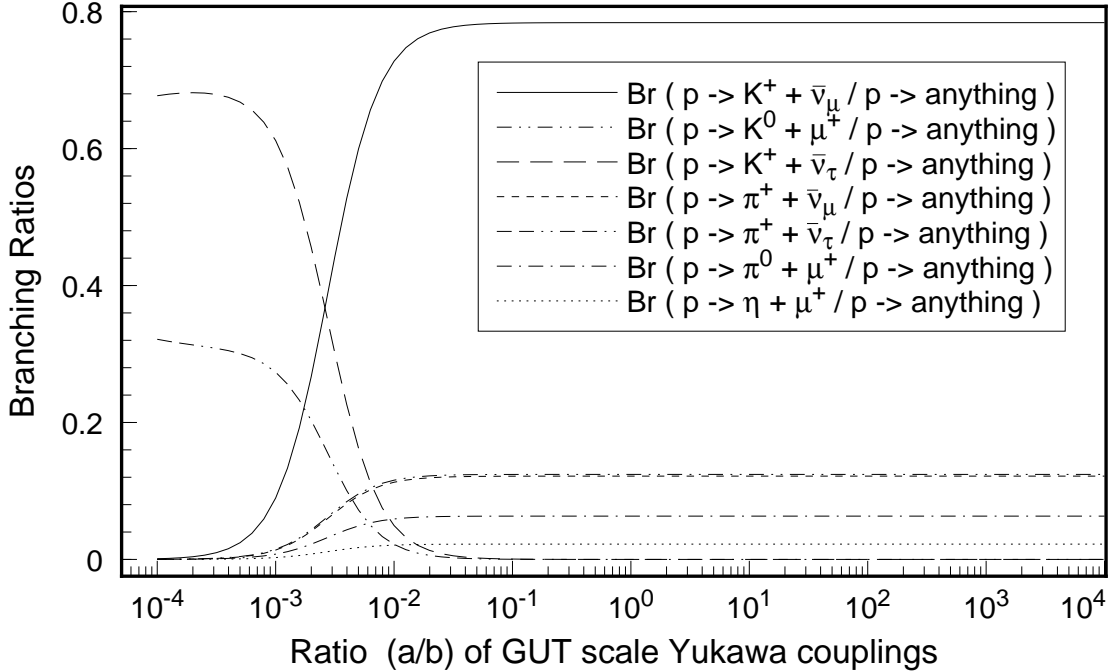}
\vspace{9.0cm}

\caption[]{The branching ratios of the most dominant proton decay channels.}

\label{figBFp}
\end{figure}

\begin{figure}[bt]
\includegraphics{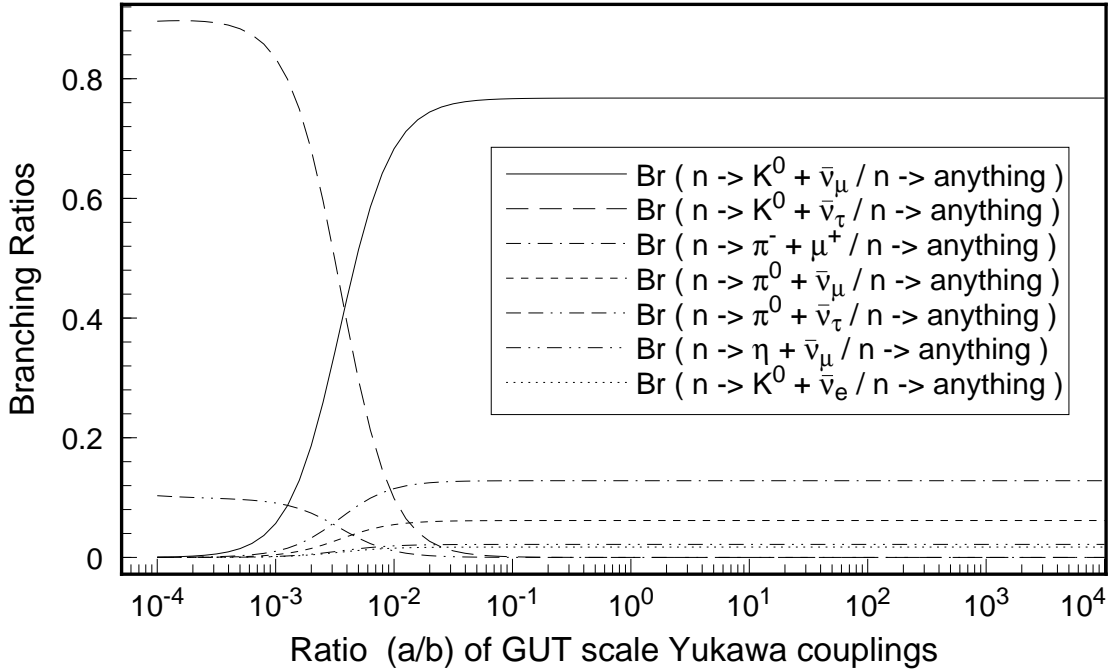}
\vspace{9.0cm}

\caption[]{The branching ratios of the most dominant neutron decay channels.}

\label{figBFn}
\end{figure}

{}From these decay rates, it is then very simple to construct branching
ratios, which have the advantage over decay rates in that most of the as
yet unspecified factors hidden in the quark level operators $C(qqql)$
divide out, leaving the branching ratios parameterised by the ratio of the
GUT scale Yukawa couplings of the complex $\underline{10}$.  The numerical
predictions for the branching ratios of the most dominant proton and
neutron decay channels, for a large range of this parameter, $ {a \over
b}$, are presented in Figures \ref{figBFp} and \ref{figBFn} respectively.
For this numerical evaluation, the branching ratios are defined as
\begin{equation}
Br(N \rightarrow x + y) = {\Gamma (N \rightarrow x + y) \over \Gamma ( N
\rightarrow \mbox{anything})}
\label{BF}
\end{equation}
where $N$ represents either the nucleon, and the decay rate for $N
\rightarrow $ anything has been taken as the sum of all the relevant decay
rates listed in (\ref{decayrates}). 

\section{Conclusions}

With the results of the analysis of nucleon decay in this non-minimal SUSY
SO(10) model presented in Figures \ref{figBFp} and \ref{figBFn}, a number
of important conclusions can be drawn. The first and most significant
point is that this model gives one-parameter predictions for all the
relevant nucleon decay branching ratios. Once nucleon decay is observed
through any two channels, the ratio ${a \over b}$ is determined, and all
the remaining partial lifetimes of the proton and the neutron then have a
definite prediction. As with the SUSY SU(5) models, this model predicts
that for a large region of ${a \over b}$ parameter space, $p \rightarrow
K^{+} + \bar{\nu}_{\mu}$ and $n \rightarrow K^{0} + \bar{\nu}_{\mu}$ are
the most dominant proton and neutron decay modes. This prediction could
only be altered by a strong suppression of the GUT scale Yukawa coupling
$a$ relative to the third family self-coupling $b$, as shown by the
prominence of the $p \rightarrow K^{+} + \bar{\nu}_{\tau}$, $p \rightarrow
\pi^{+} + \bar{\nu}_{\tau}$, $n \rightarrow K^{0} + \bar{\nu}_{\tau}$, and
$n \rightarrow \pi^{0} + \bar{\nu}_{\tau}$ decay modes for ${a \over b} <
3 \times 10^{-2}$. Another striking feature is that for ${a \over b} >
10^{-2}$ the branching ratio predictions are insensitive to the actual
value of the parameter, thereby implying a degree of robustness to the
predictions, regardless of the relative importance of the $\underline{10}$
of SO(10) in the assumed form of the GUT scale texture. 

However, it is the relative strengths of some of the individual branching
ratios that serve to identify this model, and in particular, it is the
nucleon decay channels involving the $\mu^{+}$ and the $\bar{\nu}_{\mu}$
that are the distinctive fingerprints of this model. For both the proton
and the neutron, the branching ratio predictions for channels involving
the charged muon show a marked enhancement over corresponding predictions
of minimal SUSY SU(5). Specifically, the branching ratio predictions for
the $p \rightarrow K^{0} + \mu^{+}$, $p \rightarrow \pi^{0} + \mu^{+}$,
and $n \rightarrow \pi^{-} + \mu^{+}$ relative to the dominant proton and
neutron decay channels are enhanced over the minimal SUSY SU(5)
predictions by factors of 50-500, 10-100, and 20-200 respectively (the
ranges given in these enhancement factors are due to the uncertainty of
the minimal SUSY SU(5) predictions as quoted by \cite{LL8,LL4,LL13}). To a
lesser extent, the $p \rightarrow \pi^{+} + \bar{\nu}_{\mu}$ and $n
\rightarrow \pi^{0} + \bar{\nu}_{\mu}$ decay channels show a similar
enhancement, but only by factors of 3.6 and 2.6 respectively. Thus, these
enhancements in the decay rate predictions result in branching ratios for
this non-minimal SUSY SO(10) model that are both qualitative and
quantitatively different from that of the SUSY SU(5) nucleon decay
spectrum, thereby making this `realistic' non-minimal model a testable
candidate for a SUSY GUT extension to the standard model. The issue of
testing the predictions of this model could be addressed at
Super-KAMIOKANDE, provided that Super-KAMIOKANDE in fact observes nucleon
decay. 

In sum, the distinctive tests of this realistic supersymmetric SO(10) GUT
which arise from the consideration of nucleon decay come not from the
actual decay rates or partial lifetimes of the nucleon, as the nature of
the Higgs sector and the uncertainty of the Higgs and Higgsino colour
triplet masses make the SUSY dimension 5 operator decay rate predictions
uncertain. Rather, they come from the calculation of nucleon decay
branching ratios. The fact that this realistic model predicts ratios of
branching ratios ${Br(p \rightarrow K^{0} + \mu^{+}) \over Br(p
\rightarrow K^{+} + \bar{\nu}_{\mu})}$, ${Br(p \rightarrow \pi^{0} +
\mu^{+}) \over Br(p \rightarrow K^{+} + \bar{\nu}_{\mu})}$, and ${Br(n
\rightarrow \pi^{-} + \mu^{+}) \over Br(n \rightarrow K^{0} +
\bar{\nu}_{\mu})}$ of order $20\%$, shows the relevance of `observable'
channels such as $p \rightarrow K^{0} + \mu^{+}$, $p \rightarrow \pi^{0} +
\mu^{+}$, and $n \rightarrow \pi^{-} + \mu^{+}$ to the testing of models
of GUT unification. (For related considerations involving mass textures
induced by higher dimensional operators see \cite{Murayama}.) These
enhanced branching ratio predictions are instead simply a result of the
composition of the Higgs superfield sector, which is such that the GUT
scale Yukawa couplings relevant to nucleon decay are not the full set of
couplings that contribute to SM fermion mass generation. 

The results presented here may be seen as some of the possible
implications of a viable SUSY GUT model, and any observation of $p
\rightarrow K^{0} + \mu^{+}$, $p \rightarrow \pi^{0} + \mu^{+}$ or $n
\rightarrow \pi^{-} + \mu^{+}$ at a level significantly enhanced above the
expected SUSY SU(5) predictions is an indication that the underlying
structure of a realistic extension to the standard model is best described
in terms of a SUSY GUT model with a non-minimal Higgs sector.
Unfortunately, because only a partial set of the GUT scale Yukawa
couplings is directly involved in the analysis of nucleon decay, whereas
the light Higgs is a linear combination of contributions from the various
SO(10) Higgs reps, the actual values of the GUT scale couplings remain
undetermined and the texture unexplained, at least in this model. 

\noindent{{\bf Acknowledgements}} \\ 
\noindent The author wishs to thank B.A. Campbell for suggesting the
problem and his subsequent supervision and encouragement, as well as  N.
Rodning for the useful discussions that occurred over the course of the work.


\begin{thebibliography}{99}
\bibitem{weakanom} G. t'Hooft, Phys. Rev. Lett. {\bf 37} (1976) 8.
\bibitem{Lee} T.D. Lee and C.N. Yang, Phys. Rev. {\bf 98} (1955) 101. 
\bibitem{Sakharov} A.D. Sakharov, JETP Letters {\bf 5} (1967) 24.
\bibitem{BH} S.W. Hawking, Nature {\bf 248} (1974) 30.
\bibitem{GUTsolve} For a review see P. Langacker, Phys. Rep. {\bf C72}
(1981) 185.
\bibitem{Georgi} H. Georgi and L. Glashow, Phys. Rev. Lett. {\bf 32}
(1974) 438.
\bibitem{Pati} J.C. Pati and A. Salam, Phys. Rev. {\bf D10} (1974) 275.
\bibitem{W79} S. Weinberg, Phys. Rev. Lett. {\bf 43} (1979) 1566.
\bibitem{Zee} F. Wilczek and A. Zee, Phys. Rev. Lett. {\bf B43} (1979) 1571.
\bibitem{GQW} H.Georgi, H.R. Quinn, and S. Weinberg, Phys. Rev.
Lett. {\bf 33} (1974) 451. 
\bibitem{Goldhaber} M. Goldhaber and W.J. Marciano, Comm. Nucl. Part. Phys.
{\bf 16} (1986) 23.
\bibitem{IMB} W. Gajewski, et. al., Phys. Rev. {\bf D42} (1990) 2974.
\bibitem{Barloutaud} For a review of present limits see R.
Barloutaud, {\it Proceedings of the International Workshop on
Theoretical and Phenomenological Aspects of Underground Physics (TAUP
91)}, edited by A. Morales, J. Morales, J.A. Villar (North-Holland, 1992)
p. 522.  
\bibitem{Ross} For discussions on various non-SUSY and SUSY
model predictions, refer to G.G. Ross, {\it Grand Unified Theories},
Frontiers in Physics Series (Addison-Wesley, 1985), and
R.N. Mohapatra {\it Unification and Supersymmetry} (Springer
Verlag, 1986).
\bibitem{LEP} P.Langacker and N. Polonsky, Phys. Rev. {\bf D47} (1993) 4028.
\bibitem{SUSYdim5} N. Sakai and T. Yanagida, Nucl. Phys. {\bf B197}
(1982) 553.
\bibitem{Weinberg} S. Weinberg, Phys. Rev.{\bf D26} (1982) 287.
\bibitem{GeorgiDimop} S. Dimopoulos and H. Georgi, Nucl. Phys {\bf B193}
(1981) 150.
\bibitem{Sakai} N. Sakai, Z. Phys. {\bf C11} (1982) 153.
\bibitem{DRW} S. Dimopoulos, S. Raby, and F. Wilczek, Phys. Lett. {\bf
B112} (1982) 133.
\bibitem{ENR} J. Ellis, D.V. Nanopoulos, and S. Rudaz, Nucl. Phys. 
{\bf B202} (1983) 43.
\bibitem{LL1} W. Lucha, Nucl. Phys. {\bf B221} (1983) 300.
\bibitem{LL2} V.M. Belyaev and M.I. Vysotskii, Phys. Lett. {\bf B127}
(1983) 215.
\bibitem{LL3} S. Chadha and M. Daniel, Nucl. Phys. {\bf B229} (1983) 105.
\bibitem{LL4} S.J. Brodsky, J. Ellis, J.S. Hagelin, and C. Sachrajda, Nucl. Phys. {\bf B238} (1984) 561.
\bibitem{LL5} N. Sakai, Nucl. Phys. {\bf B238} (1984) 317.
\bibitem{LL6} S. Chadha and M. Daniel, Phys. Lett. {\bf B137} (1984) 374.
\bibitem{LL7} J. Milutinovic, P. B. Pal,  and G. Senjanovic, Phys. Lett. {\bf B140} (1984) 324.
\bibitem{LL8} B.A. Campbell, J. Ellis, and D.V. Nanopoulos, Phys. Lett. {\bf B141} (1984) 229.
\bibitem{LL9} L.E. Ibanez and C. Munoz, Nucl. Phys. {\bf B245} (1984) 425.
\bibitem{LL10} S. Chadha , G.D. Coughlan , M. Daniel, and G.G. Ross, Phys. Lett. {\bf B149} (1984) 477.
\bibitem{LL11} P. Nath, A.H. Chamseddine, and R. Arnowitt, Phys. Rev. {\bf D32} (1985) 2348.
\bibitem{LL12} P. Nath and R. Arnowitt, Phys. Rev. {\bf D38} (1988) 1479.
\bibitem{LL13} R. Arnowitt and P. Nath, Phys. Rev. Lett. {\bf 69} (1992) 725.
\bibitem{LL14} J. Hisano, H. Murayama, and T. Yanagida,   Nucl. Phys. {\bf B402} (1993) 46.
\bibitem{LL15} J.L. Lopez, D.V. Nanopoulos, and H. Pois, Phys. Rev. {\bf D47} (1993) 2468.
\bibitem{LL16} K. Daum, Z. Phys. {\bf C62} (1994) 383.
\bibitem{LL17} T. Nihei and J. Arafune, Prog. Theor. Phys. {\bf 93}, 665 (1995).
\bibitem{Kamiokande}  KAMIOKANDE-II Collaboration (K.S. Hirata et.
al.), Phys. Lett. {\bf B220} (1989) 308.
\bibitem{BDO} B.A. Campbell, S. Davidson, and K.A. Olive, Nucl. Phys. {\bf
B399} (1993) 111. 
\bibitem{HRR} J.A. Harvey, D.B. Reiss, and P Ramond, Nucl. Phys {\bf
B199} (1982) 223. 
\bibitem{Fritzsch} H. Fritzsch, Phys. Lett. {\bf B70}, 437 (1977).
\bibitem{Georgi-Jarlskog} H Georgi, and C Jarlskog, Phys. Lett.
{\bf B86} (1979) 97.
\bibitem{Oakes} R.J. Oakes, Phys. Rev. {\bf D26} (1982) 1128.
\bibitem{DHR} S. Dimopoulos, L.J. Hall, and S. Raby, Phys.
Rev. {\bf D45} (1992) 4192.
\bibitem{Vcb} CLEO Collaboration, B. Barish et al., Phys. Rev. {\bf D51} (1995) 1014;\\
M. Shifman, N.G. Uraltsev, and A. Vainshtein, Phys. Rev {\bf D51}
(1995) 2217.
\bibitem{Claudson} M. Claudson, M.B. Wise, and L.J. Hall, Nucl. Phys.
{\bf B195} (1982) 297.
\bibitem{Murayama} H. Murayama, and D.B. Kaplan, Phys. Lett.
{\bf B336} (1994) 221.

\end{thebibliography}
\end{document}